Modeling theory applied; modeling instruction in university physics

Eric Brewe

Hawaii Pacific University, Kaneohe, HI 96744

ABSTRACT

The modeling theory of instruction is widely applied and highly successful in high-school instruction, and seldom in university physics. One reason is lack of familiarity with models in the physics classroom. Ongoing curriculum development has initiated application of modeling theory at the university level. This paper describes a university physics class as it progresses through a modeling cycle, including model development, application and adaptation, extension, and revision in an effort to reify the role of models in a class. Benefits of modeling instruction are identified and include effective knowledge organization and consistency with accepted scientific practice.

I. INTRODUCTION

The modeling theory of physics instruction has had, arguably, the greatest impact on high school physics instruction of any physics education reform.[1] In spite of this, it has had negligible national impact on instruction in university physics. This reality begs the question, 'Why should a valuable theory of instruction be limited to high school physics?' A number of factors have come to bear on the situation. Hestenes described modeling theory in "Toward a modeling theory of physics instruction."[2] This paper carefully laid out the elements of the theory, but it had little to do with praxis and is known to be difficult for physics professors to read.[3] A second paper, "Modeling games in the Newtonian world," used historical examples to identify elements of the theory at work, and again included little application of the theory in actual instruction.[4] The Wells, Hestenes and Swackhammer paper is most closely associated with instruction.[5] While this paper describes the activities of a hugely influential teacher, it falls short of clearly establishing the role of models and the process of modeling in the classroom. Modeling workshops, held during the summer for high school teachers, have become the best opportunity to determine the role of models in physics classes. Although the modeling workshops have been influential on the high school level, workshops have not successfully been adapted for university physics instruction. This article has two purposes, first is to reify the role of models in a university physics course by describing a class as it progresses through one modeling cycle. A second purpose is to then identify pedagogical benefits of modeling instruction in university physics classes.

A. Influence of modeling theory on modeling instruction

The modeling theory of instruction is centered on the idea that physicists reason from mental constructs known as models. Scientists begin the process of model construction by using tools such as graphs, charts, diagrams, and formulae, to represent specific physical situations. Through repeated application of the representational tools and analyses, the scientist is then able to identify general characteristics, find common patterns of use and interpretation, and coordinate the representations into a *general model* that applies to a broad class of situations. In the course of constructing situation specific models scientists accumulate experiential, declarative, and procedural knowledge that is closely associated with the model's common applications, these constitute the *modeling* component of scientists' knowledge base.[6] Examples of this type of declarative knowledge include the laws and constraints governing a model as well as a models' range of applicability and scalability.[7] The procedural and experiential knowledge associated with a model comprise the 'tricks of the trade' and make the modeling process more efficient and fruitful.

The role of models is undeniable for practicing scientists; models are the basis for research, both theoretical and experimental, which makes them the kernel for knowledge development, reasoning and problem solving.[8] However, models held by scientists are dissimilar to both the understanding students bring to introductory physics and to the standard content delivered in introductory physics. Students' comprehension of the physical world at the beginning of introductory physics is a fragmented collection of common sense generalizations which are primarily pre-scientific.[9] The content organization in introductory physics does not help students develop model-centered knowledge bases, in that it is a litany of isolated topics. Textbooks are the worst offenders in this regard; they attempt to deliver vast quantities of declarative content, but do not attend to the procedural elements that make the content practical. Modeling instruction was designed to help students develop model-centered knowledge bases that resemble those of practicing scientists.

### B. Content organization in modeling instruction

In the modeling theory of instruction, a primary role of models is to simplify the content of the introductory course. Modeling instruction organizes the content of introductory physics around a small number of *general models* that can be applied in a broad array of situations. Two benefits are derived from focusing the curriculum on 6-8 general models. First, the curriculum organization matches expert knowledge organization. Second, students see a small number of general models as a manageable body of knowledge, whereas the current organization of the introductory curriculum is untenable. Table I summarizes the differences in content organization between model-centered content and standard content.

Table I. Comparison of model-centered content and standard content for introductory physics.

| Model-centered content | Standard content |
|---|---|
| Models are constructs that are built in accordance with physical laws and constraints | Laws are given, in equation form, and applied to solve problems |
| Models are built through application of representational tools which can then be used to solve problems | Problem solving is primarily quantitative manipulation of equations |
| Models are temporal, they must be validated, refined and applied | Content is permanent, all validation has already taken place |
| General models are applied to specific physical situations | Laws apply to specific situations |
| Modeling is a process that is learned through accumulating experience | Problem solving is a game that requires tricks and is learned by solving large numbers of problems |
| Models are distinct from the phenomena they represent and can include causal, descriptive and predictive elements | Content is indistinguishable from the phenomena |

C. Pedagogy in modeling instruction

Modeling instruction not only focuses the content on a small set of general models, but also influences pedagogy. Modeling instruction is based on a theory of science. Extrapolating the theory of science to a theory of instruction, you arrive at the expectation that both the curriculum organization and the pedagogy foster scientific behavior from students. Modeling instruction primarily endeavors to have students engage in activities that are consistent with the activities of practicing scientists. Students are treated as neophyte physicists, learning the practice of physics. The outcome is that modeling utilizes an inquiry-based approach, and is ideally suited to studio-format classes. This differs from the lecture/lab/recitation method in that artificial separations between experimental and theoretical considerations are removed with studio-format classes. The inquiry approach of modeling instruction focuses on developing an understanding of physical situations by creating models of physical situations.

### D. Construction of models

Central to both the pedagogy and content organization in modeling instruction is the process of model construction. However, the mechanics of model construction are not well known. Halloun describes it as a "middle-out" process, meaning that students examine phenomena by building models and that models exist on a structural level below the laws of physics and above the level of individual concepts.[10] Using this description, the outcome of model construction is that students are able to use models to explain broad classes of individual concepts as well as examine the behavior of models as dictated by laws of physics.

To clarify the mechanics of model development, I will describe the instruction that leads to the general constant acceleration model. Students begin to construct general models by first learning the representational tools and building up experiential, declarative, and procedural knowledge. Accordingly, the instruction begins with a phenomenological introduction through inquiry-based lab activities. For constant acceleration, the lab includes students moving in front of motion detectors and interpreting the kinematic graphs that result. Additionally, students begin accumulating declarative knowledge in the introductory inquiry labs by identifying the important concepts (position, velocity, distance, displacement, speed, acceleration) and developing working definitions. After students have been introduced to the representational tools, in this situation the kinematic graphs, the instruction turns to coordinating multiple representations. In the case of constant acceleration, students would engage in conceptual activities such as interpreting graphs (slope and areas under v vs. t) creating corresponding motion maps, and vice versa. Learning the representational tools and beginning to coordinate multiple representations happens in the first two phases described in Table II.

Table II. Modeling instructional cycle leading to development of general constant acceleration model.

| Model construction | Instructional goal | Student activity |
|---|---|---|
| Introduction and Representation | Phenomenology –initiates the need for a new model (accelerated motion is not accounted for by general constant velocity model.) Introduction of kinematic graphs as useful representation. | Experimentation involving students moving with constant acceleration in front of motion detectors. |
| Coordination of Representations | Relate the kinematic graphs to other common representations (motion maps). | Experimentation and conceptual activities |
| Introductory application | Extend application | Develop kinematic equations from kinematic graphs by analyzing velocity vs. time graphs. |
| Application | Develop experience, heuristics, and learn to draw conclusions based on representations. | Problem solving emphasizing use of modeling tools. |
| Abstraction and Generalization | Identify characteristics of representations in situations involving constant acceleration. | Review of constant acceleration and guided discussion. |
| Continued Incremental Development | Relate constant acceleration model to dynamical models and apply to new situations | Continually revisit constant acceleration model, coordinate with energy and forces, apply to electricity and magnetism. |

Along with the introduction of each representation, there are applications, or opportunities for students to use their new tools.  In traditional instruction this would correspond to the problem solving element of the class. The nature of the problem solving situations and the review of the situations distinguish modeling instruction, which I will describe in more detail later.  Within the application phases of the instructional cycle, the situations are chosen such that students are applying the tools they have learned to model specific situations, by doing this the students are building up experiential, declarative, conceptual, and procedural knowledge.  Technically, during the application phases, students are modeling specific situations even though the general constant acceleration model has not been developed.

After sufficient applications, the instructor leads a series of class discussions to help students organize their experiential, procedural, and declarative knowledge into a general model.  For an example of the general characteristics of a model, see Table III, which describes the general characteristics of the constant acceleration model.  In this series of discussions students call on their experience in analyzing physical situations and compare situations for similar characteristics in the application of representations.  The definitions, graphs, equations, and interpretations from the situation specific models are then collected and a whiteboard meeting is used to generalize the characteristics of all constant acceleration situations into a single, *general,* constant acceleration model. This group of representations, experiential, declarative and procedural knowledge evolves into a general model when they are collected and abstracted into a single, coherently-organized unit.  By abstracting into a general model students are constructing a general model that is significantly different than the situation specific models they have been constructing. Until this point, students have been modeling specific physical situations; the general model is different because the abstraction requires students to look at characteristics that are common to all constant acceleration models. Cognitively, generalization is critical because it groups representations to reduce the cognitive load on the student by allowing recall of a single general model rather than an array of distinct situation specific models, which in turn simplifies the curriculum. Admittedly, the unification of the representations, experiential, declarative, and procedural knowledge elements into a single model does not happen with one discussion, but instead incrementally over the course of a semester and requires maintenance on the part of the instructor.

Once the generalization of a model has taken place, the general model becomes a template which can be applied and improved through repeated application in new contexts. The continued incremental development of the general constant acceleration model begins with applications to 2-d motion, which, in the minds of the students, is a situation where it no longer applies. Applying the general model in a new context often requires a revisiting of the first four stages of the modeling cycle identified in Table II, although the introduction and representation and coordination of representations are greatly facilitated by relying on the students understanding of the general constant acceleration model. This process begins with students examining an object undergoing 2-dimensional motion. The instructor then introduces vector diagrams and vector mathematics which extend the utility of the constant acceleration model to include two-dimensional motion.[11] Again, the constant acceleration model is applied to two-dimensional motion until it no longer adds to students' procedural knowledge.

The continued incremental development of the model begins, but does not end, with 2-d motion. Instead the constant acceleration model will be used repeatedly throughout the instruction, which I will subsequently describe. The development of the general constant acceleration model is about building a single knowledge structure that can explain broad classes of phenomena. The ongoing incremental development will be about the behavior of models according to the laws and theories of physics.

Table III. Characteristics of a generalized constant acceleration model

| Generalized Constant Acceleration Model | |
|---|---|
| Kinematic Graphs (For 1-d motion) | Position vs. time graphs are parabolic<br>    Slope of tangent = instantaneous velocity<br>Velocity vs. time graphs are linear<br>    Bounded area = displacement<br>    Slope = acceleration<br>Acceleration vs. time are horizontal<br>    Bounded area = $\Delta v$ |
| Motion Maps | Velocity vectors are constantly changing<br>Vector subtraction gives direction of acceleration |
| Energy Pie Charts | Kinetic energy is constantly changing |
| Force Diagrams | Net force vector is non zero |
| Kinematic Equations (Valid as vector equations) | $\mathbf{v_f} = \mathbf{v_0} + \mathbf{a}t$<br>$\mathbf{d} = \mathbf{v_0}t + \tfrac{1}{2}\mathbf{a}t^2$ |

## II. MODEL USE IN INSTRUCTION

To this point, I have described instruction that leads to the development of a general model, the constant acceleration model. Currently, I will describe the instructional uses of a general model. Because the general model continues to be deployed and applied, the general model will continue to develop incrementally in an ongoing cycle of application and development.

A. Context

In order to reify the utility of models in a university physics course, it is imperative to examine a part of a course. In this case, the part of the course described will be the introduction of energy, which immediately followed 1-d and 2-d kinematics, and preceded the introduction of forces. The course I describe is an algebra-based course that meets twice a week in three hour meetings, so the description runs five class meetings over 2 ½ weeks. This particular class has 29 students, which is ideal for a student-centered studio course.[12] Table IV identifies the timeline and activities that are described.

Table IV. Timeline of activities in the energy introduction unit of a modeling course

| Course meeting | Activity | Intent | Topic |
| --- | --- | --- | --- |
| Day 1 | Ball bounce | Model ramification Tool introduction | Introduction of energy conservation-qualitative |
| Day 2 | Quantitative energy lab | Model extension Model adaptation | Application of energy conservation-quantitative |
| Day 3-4 | Modeling physical situations | Model application Model adaptation | Energy problem solving |
| Day 5 | Modeling static situation | Model extension | Introduction of forces |

B. Shifting from descriptive to causal models

Once students have generalized the constant acceleration model for both 1-d and 2-d kinematics, the model is extended when the instructor presents a situation that requires the inclusion of energy. Nothing in the models to this point has been causal.

Students begin their introduction to energy with a lab, which builds on their kinematic models. The lab first requires the students to create descriptive models of the motion of a ball from the time it is dropped until it reaches its highest point after the first bounce. They are able to use the general constant acceleration model as a template to model this motion. Then in small lab groups, students collect data to validate (or invalidate) the predictions they generated with their kinematic models. While the students are working with the computers to collect data, the instructor asks one or two select lab groups why the ball doesn't return to the original height. This type of instruction is predicated on the instructor having experience that students will generally respond "energy is lost". Once energy has been introduced *by a student*, the instructor continues by asking what is 'known' about energy; predictably, students' responses are nearly uniform "energy is neither created nor destroyed" like a mantra. The instructor then engages the selected lab group in a discussion of energy conservation using common sense questions such as, "If the ball has energy at the bottom, and energy is conserved, that energy must have come from somewhere. Where could it have come from?" The instructor then uses the need to track the storage and transfer of energy to introduce energy pie charts. Energy pie charts, which are an adaptation of Van Huevelen's energy bar charts, are used because they allow for qualitative analysis of the ball bounce situation.[13] The selected lab group then is given the responsibility to introduce the representational tool to the rest of the class during a whiteboard discussion session, or "board meeting". In this board meeting one of the essential discussion points is the relationship between the existing kinematic models and the newly introduced representational tool. This discussion is critical because it validates the representational tool, and ensures self-coherence within the model. Students have used the models that they already know to investigate a phenomenon that is unknown. They have been introduced to a new representation that can be incorporated into their kinematic models, and because it provides causal explanatory power, it improves the power of their models. In this manner students have begun to make incremental improvements to the general constant acceleration model by seeing how the model behaves in accordance with conservation of energy.

C. Extending the model, becoming quantitative

Modeling instruction relies on various representational tools because they enhance students' conceptual reasoning about physical situations. Prior to spending significant time calculating energy before and after some event, students in a modeling course would do one homework assignment and have at least one board meeting that both relate to the use of energy pie charts and their interpretation. These activities help the students develop qualitative understanding of energy without adding mathematic complexity. Although analysis and prediction are possible using only energy pie charts, the applicability is limited. Clearly, students need quantitative tools for energy.

The development of the equations, $E_k = ½ mv^2$ and $E_g = mgh$, is guided by comparing the models developed for three situations. The three situations are included in Table V. Students construct models for these three situations as homework, and begin class with a discussion of the models. The models students create include kinematic graphs and equations as well as energy pie charts. The instructor leads students to compare the models, and from these differences, infer which variables affect $E_g$. The model for Situation #1 is used as a baseline, and shows energy transferring from $E_g$ at the top to $E_k$ at the bottom. Situation #2 looks similar, but because the book has additional mass, they infer that the energy pie charts for #2 would be larger than for #1. This leads to discussion about the energy required to lift a heavy book as compared with a light book which relates the representation to students' experience. Once students recognize the size of the pie charts is important, they are able to determine that Situation #3 also has larger pie charts than #1. The instructor would steer the discussion to the variables that are related to $E_g$ and $E_k$. By comparing the set of pie charts for Situations #2 and #3 to Situation #1, students come to consensus that $E_g$ is related to mass and height. At this point the instructor gives the equation, $E_g = mgh$. But the foundation for this equation is rooted in the models they have previously made of the situation. New information is introduced as it becomes useful.

From there, students are challenged to propose experiments that will allow them to find the equation for $E_k$. The proposals students generate are required to use the models created as the basis for proposing and conducting experiments. Students then carry out the experiments they proposed and data are analyzed to validate the equations for kinetic and gravitational energy. Students are assigned homework to use data to validate the equation for $E_k$.

Table V. Three modeling situations used in the quantitative introduction of energy.

| Situation #1 | Situation #2 | Situation #3 |
|---|---|---|
| A 3 kg physics book is dropped from a height of 0.1 m. | A 5 kg physics book is dropped from a height of 0.1 m. | A 3 kg physics book is dropped from a height of 0.5 m. |

The introduction of energy in modeling instruction differs from traditional curricula in three important ways. 1. Energy is introduced prior to forces, and is always used in the context of energy conservation. Traditional curricula tend to introduce work, and the work-kinetic energy theorem, which is consistent with a force-centered approach to the content and has negative theoretical implications.[14]  2. Energy is used to extend the kinematic models that students have developed in the first part of the course.  New content is introduced to extend the applicability of models.  Rather than in the standard approach where curriculum is organized such that new content is introduced in distinct chapters, which atomizes the curriculum, and leads to students missing the coherence of the subject.  3.  The various representations that comprise the model (kinematic graphs, motion maps, energy pie charts, and the associated mathematics) are the focus for the students; they are not asked to rely solely on mathematical representations.  Multiple representations of physical situations are a central element in modeling instruction; therefore, the representations are not just introduced, but students are expected to solve problems by utilizing the representations that make up models.

### D. Model application and adaptation

The third day of class begins with a whole class discussion of the equations generated for $E_k$ and students show data to support the equations.  Validation of the new elements in the model is essential because it allows students to feel confident in adapting their general models to accommodate the quantitative representation for energy.  Once students have developed the quantitative representations for energy, they then practice using these representations by modeling physical situations, this is the model adaptation and application phase.  Over the remainder of days three and four, students are engaged in creating models of physical situations that now include quantitative modeling of energy storage and transfer.  The first physical situation presented to the students is a situation they have already encountered while creating 1-d constant acceleration models, but this time they are instructed to include energy.  Using a situation they have already modeled provides an outlet to discuss how the models change with new elements, how the models have to be adapted, and how they are more powerful, useful and efficient.  Students then apply the models to additional situations by creating a model for a specific situation and presenting the model on a whiteboard.  Whiteboards are discussed by the whole class and during these discussions the instructor highlights exemplary models and procedural pitfalls.

      While this part of the class is analogous to traditional problem solving it should not be considered equivalent.  The application and adaptation phase of the modeling cycle has major philosophical differences from traditional curricula, as traditional curricula primarily assess students on well defined physics problems with clear questions.  The difference in philosophy here presents a difficulty in getting students to view models as the goal rather than specific numeric answers.  In order to clarify the situation, I have created a comparison between a standard textbook problem and how the same problem would be used in a model-centered course.  The example in Figure 1 is a problem taken from Understanding Physics[15] which was modified to be consistent with a modeling approach.

| Standard Problem Statement | Modeling Problem Statement |
|---|---|
| **Block Drawn by Rope:** A 3.57 kg block is drawn at constant speed 4.06 m along a horizontal floor by a rope. The force on the block from the rope has a magnitude of 7.68 N and is directed 15.0° above the horizontal. What are (a) the work done by the rope's force, (b) the increase in thermal energy of the block floor system, and (c) the coefficient of kinetic friction between the block and floor? | Construct a complete Model of the following situation: A 3.57 kg block is drawn at constant speed 4.06 m along a horizontal floor by a rope. The force on the block from the rope has a magnitude of 7.68 N and is directed 15.0° above the horizontal. |

Figure 1. Comparison of problem statements from standard textbook problem and Modeling problem. (Problem #41, Chapter 10, p 290 Understanding Physics, Cummings, Laws, Reddish and Cooney)

Student responses to these two different problems will vary greatly. The standard problem will have responses that are numeric answers and will be accompanied by varying degrees of work and likely little justification on how, or why the answer was attained. The response to the modeling problem is a constant velocity model, adapted to the situation described. A complete model for this situation would include kinematic graphs, motion maps, a system schema, a force diagram, and energy pie charts, as well as applications of Newton's Second Law and the First Law of Thermodynamics. All of the information asked for in the standard problem should be available by interpreting the model. The model answer would be much richer in representation and would be easy to troubleshoot in analysis. Often, when students create rich models, they are able to identify problems within their reasoning, or to validate their answers through redundancy within the model.

In order to encourage students to see the value in creating a model, the problems must be chosen so that the answer is a model. However, it takes significantly more time to create a rich model than to answer a standard physics problem, as a result, the number of problems assigned must be much smaller than in a standard course.  By assigning smaller numbers of problems (on the order of 2-3/week) students see that the emphasis is on quality and richness of the model rather than right answers.  By using a small number of carefully chosen problems, students can efficiently learn the procedural aspects of modeling in each given area because a well constructed model solves a large number of problems all at once, thus requiring fewer assigned problems. Also, the grading must change in a modeling course, the grade must reflect whether or not the student actually created a model, rather than whether an answer is achieved.  In some ways this is the most challenging for instructors new to modeling.

### E. When a model reaches its limits

Model application and adaptation is complete when students are no longer adding new procedural knowledge.  Instead of abruptly stopping one topic and moving to the next, in modeling instruction, the next topic is introduced when students' existing general models are insufficient to explain new phenomena.  In this class, students are able to use kinematics and energy in the construction of models, but have not yet reached forces.  When it is time to introduce a new topic, students are asked to model a situation where their models break down.  In this case it is very simple; the instructor chooses a static situation.  There is no motion and no energy transfer, so the students' models lack any content.  The only option is then to move on and begin the cycle over, with experimentation, representation, and application.  It is important to note that as the cycle starts over students are able to clearly establish one criterion for choosing one modeling approach over another. Energy approaches are useful when energy transfers are taking place, but forces are the appropriate approach when modeling static cases.

## IV. EXAMINATION OF MODELING INSTRUCTION

### A. Differences between modeling and traditional curricula

Now that I have described one modeling cycle, it is useful to consider how the curriculum must change to accommodate modeling. Halloun[16] describes differences in organization between traditional and modeling curricula. The primary difference is the traditional curriculum is organized into discreet topics, while the modeling curriculum is organized around general models that continue to build on themselves. Differences in curricular organization manifest into the primary difference between modeling instruction and traditional instruction, which is that the model-centered organization encourages cyclic reexamination of the content through model revision and ramification. Model-centered organization is critical because it simplifies the content in the mind of the learner and allows students to organize their knowledge around a small number of models rather than a large number of seemingly unrelated topics. As Reif and Heller asserted, this knowledge organization is essential for optimum problem solving performance.[17] The modeling theory of instruction extends Reif and Hellers' premise by saying the design of the introductory curriculum must mimic the organization of expert physicists' knowledge base.

### B. Model-centered curriculum

The coherence of physics is unmatched in other sciences. This elegance, coherence, and simplicity has been lost in standard curricula, likely due to the need of physicists need to break problems down into small units. The standard organization of physics curricula atomizes the content of the introductory course and students fall victim. They view the content as an extensive collection of unrelated topics, and are overwhelmed by the volume of material covered. The model-centered curriculum is designed to address this issue, by structuring the curriculum around a small number of general models that apply broadly. This organization helps students manage their knowledge base by simplifying the content into the 6-8 general models, the laws that govern models, and the procedural rules which establish how models are constructed and interpreted. This organization is more efficient because all of these elements work together, rather than a traditional curriculum where the application of laws and theories are not coherently applied beyond the scope of the chapter where they are introduced. The unification of representations into models also helps to reduce the cognitive load on students during the analysis of physical situations.

## C. Modeling as a theory of science

A second benefit of modeling instruction is the relationship between the curriculum design and the practice of science. Because modeling theory is a theory of science, the derivative, modeling instruction, provides students with a learning experience that is representative of the work of scientists. Modeling is, by nature, iterative and the iteration is built into the curriculum and is explicit. As I have demonstrated, students in a modeling cycle begin with a model that has limited applicability. Through experimentation and observation, the model is revised to incorporate new information. The new information is applied and procedural knowledge is developed through application. The cycle ends as it begins, with the model breaking down and requiring further revision.

Additionally, models represent a more plastic version of scientific knowledge. Because models have limitations, which are regularly examined, students begin to see scientific knowledge as a work in progress. Traditional curricula rarely specify the underlying assumptions, limitations, or range of applicability when presenting new content, whereas these elements are essential to models. Traditional curricula foster understanding of physics that is inconsistent with the physicists' view of the discipline, that certain models apply in certain contexts and no model is absolute.

### D. Modeling and Problem Solving

The third major benefit of the modeling approach is that it requires use of multiple representations. Many studies have identified the benefits of multiple representations in problem solving[18,19,20,21]. By making the representations an essential element of the course content, this provides students an array of powerful tools. An extensive repertoire of tools is important because they allow students to have varied ways of analyzing physical situations, including qualitative or quantitative diagrams, graphs, or equations. While many reform physics courses are beginning to include increased emphasis on representations, traditional courses almost exclusively rely on mathematical representations in solving problems. Because the modeling approach is systematic about the use and analysis of representations, it encourages students to attend to conceptual aspects in the analysis of physical situations. Larkin et. al.[22] related these conceptual analyses to greater success in problem solving. Brewe has shown evidence of improved problem solving for students in the modeling curriculum on energy problems[23].

### E. Impediments to adoption of modeling instruction

Although there are many benefits to modeling instruction, there are a number of impediments to widespread adoption. One impediment is the lack of understanding of the role of models in instruction, which I have clarified through this paper. This is not the only difficulty with adopting modeling, however, as there are not significant resources available to facilitate adoption at the university level. The high-school modeling curriculum is valuable, but is inadequate for a university-level modeling course. With the exception of Chabay and Sherwood's Matter and Interaction series[24], textbooks actually hinder instructors' abilities to run a model-centered physics course by ignoring the role of models and omitting extensive use of multiple representations. Additionally, modeling instructors must have a bank of carefully designed activities and have a working understanding of students' existing knowledge as it evolves throughout the course. Currently, a complete set of such activities does not exist. These elements compound to impede the adoption of modeling instruction.

Finally, as with any reformed curricula, the argument can be made that it requires more time. However, while the content coverage is not as great, evidence of the success of modeling at the university level has shown greater conceptual gains than traditional instruction. Modeling rests on students learning and using multiple representations. The first semester might well be considered a time to learn the tools of the trade, and the second semester an application and extension of these tools. The investment of time required to teach the representational tools early in the course is significant, still, the case can be made that proficiency with these tools early in the course leads to greater efficiency later in the course. I suggest exactly this in regards to momentum conservation. Students in a modeling course take less time to understand momentum conservation than traditional students because energy conservation has been a consistent theme for them throughout the course.[25]

## V. CONCLUSION

Modeling has greatly impacted high-school physics instruction, and yet has not impacted university physics. The lack of resources at the university level is one reason. Compounding this problem has been that the benefits of model-centered instruction have been unclear, thereby making it untenable for most instructors to adopt modeling instruction. In this paper, I have described a class as it progressed through a cycle of model revision, development, application and a return to further revision. In doing so, I have made the role of models more tangible and identified the benefits derived from modeling instruction.

## ACKNOWLEDGEMENTS

I would like to thank Charles Henderson, Leon Hsu, and Bruce Birkett for helping initiate this work. Preparation of this manuscript was supported by a National Science Foundation CCLI grant (DUE #0411344).